# Electron Ion Collider: Physics and Prospects

**Hugh E. Montgomery**[1]
*Jefferson Lab*
*12000 Jefferson Avenue, Newport News, VA 23606, USA*
*E-mail:* `mont@jlab.org`

We give an outline of the anticipated physics program for a future electron ion collider. The status and prospects for construction of such a device are discussed.

---

[1]Speaker

1. **Deep Inelastic Scattering and Parton Structure in the next Decade**

There continues to be considerable interest in deep inelastic scattering, the parton structure of the nucleus and the nucleon, and hadron physics more generally. The facilities which we expect to be operating during the next decade include Compass [1] at CERN, the 12 GeV CEBAF [2] at Jefferson Lab, and RHIC [3]. On a smaller scale we will see contributions from the Drell-Yan program exemplified by SeaQuest [4] at Fermilab, and the smaller electron accelerator facilities at Mainz and Bonn.

These facilities will make considerable progress and extend our understanding in a number of areas: meson and baryon spectroscopy, the proton charge radius, the nucleon valence structure, the form factors, the spin distributions. We can expect big strides in the tomographic studies using transverse momentum distributions and generalized parton distributions, which could lead to important insight into the role of the parton orbital angular momenta. Some progress can also be anticipated in understanding the role of the gluon in the spin of the nucleon, and the sea quark distributions. Not to be neglected are the contributions to the broader aspects of nuclear and particle physics, the searches for heavy photons, and the neutron distributions in several nuclei. We might expect that a decade from now, our ignorance will be dominated by the gluons.

2. **Need for an Electron Ion Collider**

Since ~1970 we have known that gluons must carry 50% of the momentum of the nucleon, this fact is the origin of the gluon concept. Massless gluons and almost massless quarks, through their interactions, seem to generate more than 98% of the mass of the nucleons. Without gluons, there would be no nucleons, no atomic nuclei,… no visible world!

We do not know, but suspect that the gluons carry a finite fraction of the spin of the nucleon. We believe that the residual component of the gluon interaction, a "strong interaction van der Waals" force, is the nucleon-nucleon force which controls the internal structure of the nucleus. Our present knowledge of the gluon distribution suggests a divergence as $x_{Bj}$ becomes ever smaller, see Figure 1. However the non-Abelian nature of QCD implies a saturation when the gluon recombination probability matches the gluon splitting. This saturation has not yet been clearly seen experimentally. With low $x_{Bj}$ and nuclear targets we can expect to observe this effect.

The scenario sketched above has motivated the international nuclear physics community to map out a path to elucidate the kinematic regime at low $x_{Bj}$. The ideas have crystalized into a vision of a high luminosity electron collider with well controlled spin of both electron and hadron beams at moderate and flexible energy ad a full suite of nuclear beams. The accessible region of the nucleon would range from $x_{Bj}$~0.1 to $x_{Bj}$~0.0001.

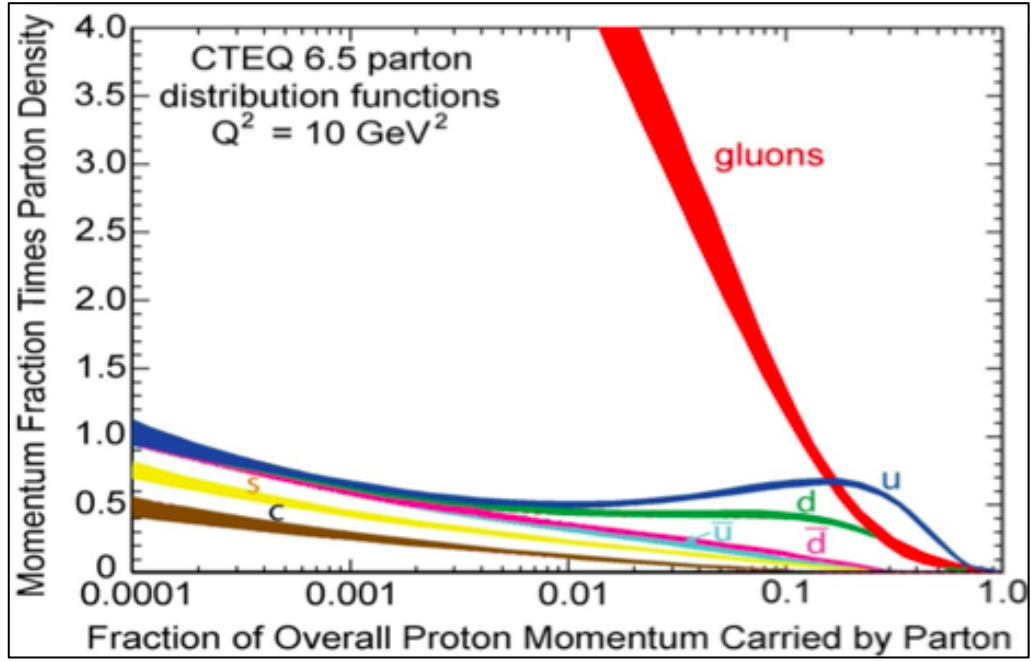

**Figure 1**: Parton Distributions as a function of $x_{Bj}$ showing steep rise of gluon distribution at small $x_{Bj}$.

### 3.    Physics with the Electron Ion Collider [5]

When we look at the plots of the F2 structure function for a hydrogen target with data from HERA and from the fixed target muon scattering experiments at CERN and Fermilab, one is tempted to conclude that our knowledge is fairly complete. This is a false impression. For the unpolarized structure functions we still lack good data, at very high and very low $x_{Bj}$ from a neutron target which limits our knowledge of up and down quarks separately. The polarized structure function data are also very sparse and very limited in kinematic range. Some of this will be rectified by the 12 GeV program at Jefferson Lab, particularly in the high $x_{Bj}$ region but the potential improvement from an electron ion collider with polarized beams is enormous.

The preceding commentary with respect to the nucleon structure functions holds even more dramatically for nuclear targets. Our understanding, for example, of the gluon distributions in heavy nuclei is almost absent. For the sea quarks, the low $x_{Bj}$ region is especially lacking. For the spin structure functions the goals are rather clear. We need to clarify the origin of the spin of the nucleon. We know that the quark contribution, per se, does not suffice. There are indications, from the RHIC data that the gluon has a finite role to play, but there is ample room for contributions, perhaps dominant, from other sources, such as the parton angular momentum. The electron ion collider will offer a definitive determination of the gluon contributions, see Figure 2, dramatically reducing the uncertainties left by the RHIC data.

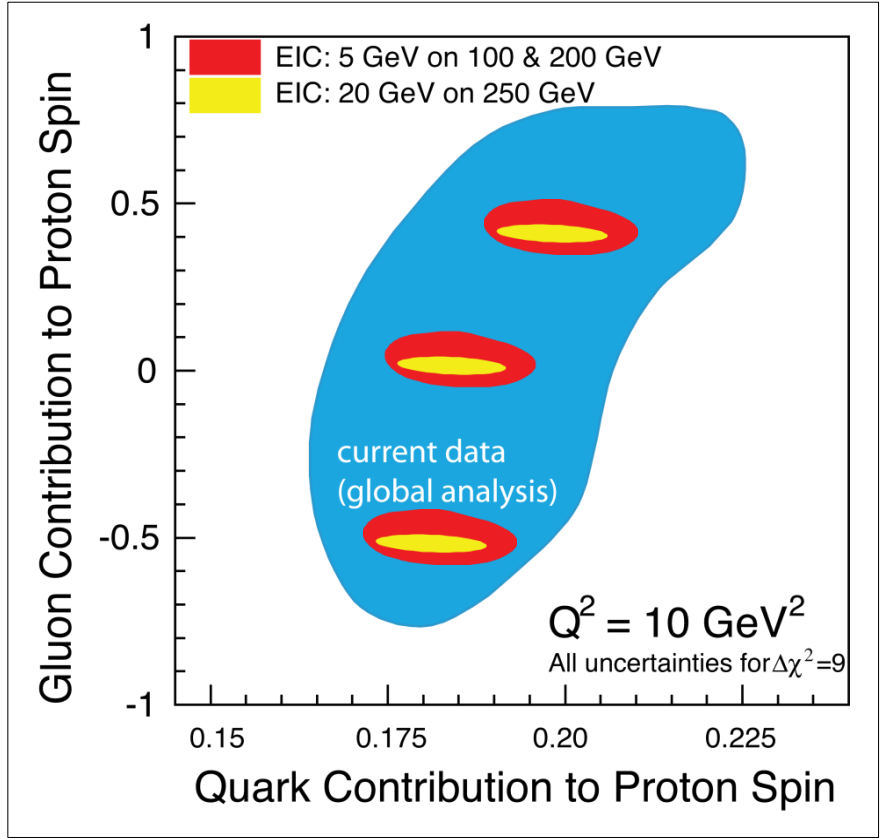

**Figure 2:** Determination of the contribution of the spin og the gluon to that of the nucleon as projected to be measured by an electron ion collider [6].

### 4.     Nucleon Tomography

For five decades we have sought to understand how charge and magnetism are distributed in the nucleon or the nucleus. Since the 1970's we have related these form factors, and the longitudinal momentum, $x_{Bj}$ distributions to the disposition of partons, quarks and gluons within the target. Since the early '90s the appreciation of these distributions has deepened as we have related them to an underlying 5-dimensional quantum mechanical phase space of the partons. Today, we are at the threshold of an era in which we probe this phase space in 3-dimensions of the momentum space by measuring the transverse momentum distributions (TMD). In these measurements we integrate over the spatial distributions. Analogously, by integrating over the transverse momentum components, the Generalized Parton Distributions (GPD) measured in Deeply Virtual Compton Scattering provide a multi-dimensional spatial view of the nucleon. Thus, by dint of improved understanding of what we are doing, and by improved experimental techniques we hope to open a whole new vista of the micro-world.

### 5.     Saturation

Our initial understanding of the evolution with momentum transfer squared of the parton distributions involved the splitting functions as the quarks and gluons radiated gluons, the DGLAP equations. Very dramatically this is seen at low $x_{Bj}$ where the gluon distribution rises steeply. However the non-abelian nature of the interactions implies also a recombination, particularly among the gluons. Thus the observed distributions should arise from a competition between radiation and this recombination; perhaps the distribution reaches a certain value and remains the same as $x_{Bj}$ changes. That is to say, maybe there is apparent saturation of the gluon distribution. This is a key signature of QCD as we understand it. Thus far it has not been observed. In principle the ratio between the diffractive and non-diffractive components of the cross-section provides a direct handle on the saturation phenomenon, as illustrated in Figure 3. At HERA diffraction accounted for about 10-15% of the cross-section. If at the electron ion collider we are seeing saturation and the emergence of a color glass condensate, that ratio would be 25-30%. There are other possible experimental signatures, but his would be the most direct.

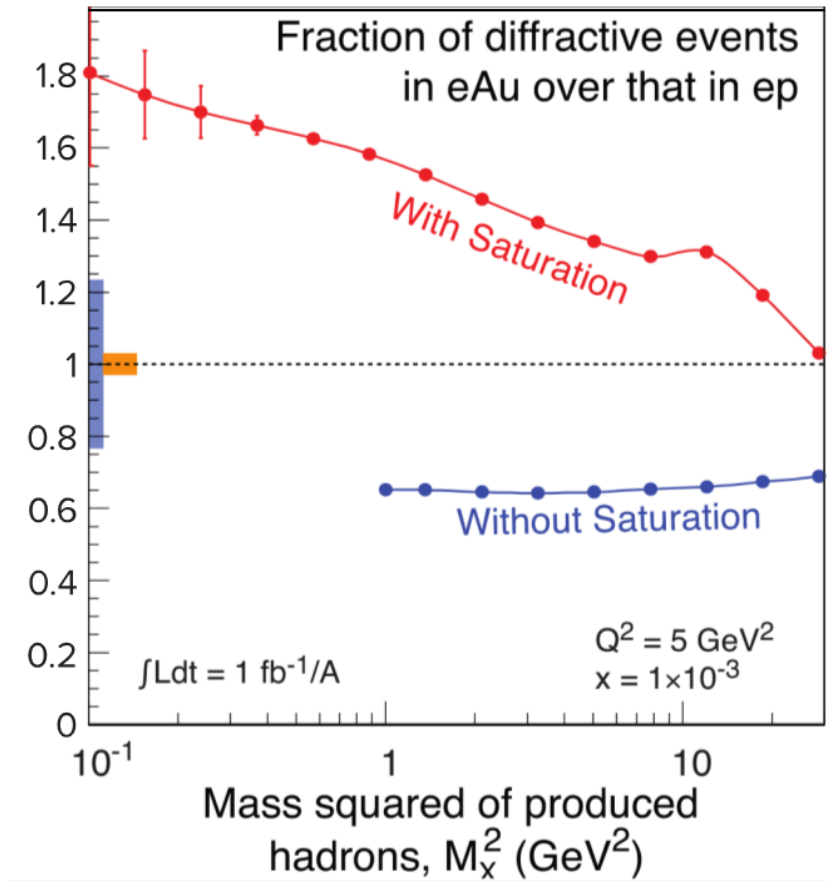

**Figure 3:** Fraction of events which are diffractive, as function of mass squared of produced hadrons, with and without gluon saturation in electron-gold collisions [5].

## 6.     Partons to Hadrons

We describe the hard underlying reactions in terms of bare partons. What we actually detect are hadrons. We would like to explore this transition experimentally. In a sense this is the exploration of the way confinement occurs. Thoughts about this process usually have involved the concept of a formation length or a formation time. By adjusting the energy, time dilation allows us to elongate the formation length in the frame of the nucleus. Thus we can consider the propagation of the initial or intermediate state through the nucleus and examine the effects, if we should see further interactions resulting in more particles with lower momenta or possibly transverse momentum broadening. Using different nuclei we can also change the length this nuclear target presents to the intermediate or final state. Studies along these lines have been tried [7, 8] with inconclusive results. An electron ion collider would represent the experimental tool par excellence for such studies. We could explore whether heavier quark states behave differently from light quarks? We could also imagine attempts to make contact with the data in heavy ion collisions, in which we imagine that the intermediate, and eventually final hadronic states are propagating through hot nuclear matter rather than the cold nuclear matter in the electron ion collider.

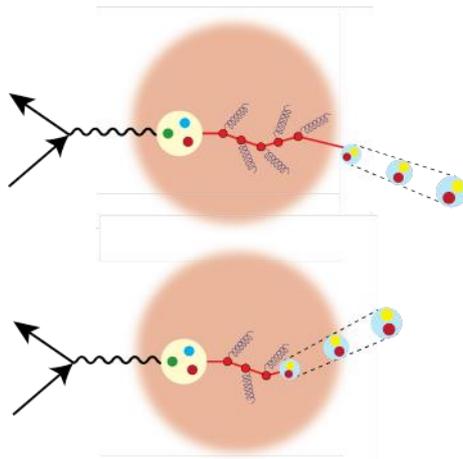

**Figure 4:** Notional sketch of the propagation of the intermediate states of hadronization through nuclear matter at two different energies.

## 7. The Collider

The community white paper advocating the construction of an electron ion collider embraces a set of performance parameters, which would satisfy the physics goals therein. In particular, the accelerator/collider complex should support collisions of electrons with the full spectrum of nuclei in which the beams of the lighter elements should be polarized. The collider center of mass energy should be flexible between 20 and 100 GeV, thereby enabling $x_{Bj}$ ~0.0001. The electron-proton luminosity should be in the range $10^{33}$- $10^{34}$ cm.$^{-2}$ s.$^{-1}$. Ion beams up to uranium should be possible.

The designs at Brookhaven National Laboratory (BNL) [9] and Thomas Jefferson National Accelerator Laboratory (Jefferson Lab) [10] build upon the existing infrastructure at the two laboratories.

At BNL, one of the existing RHIC collider rings would be used for the ion beams, hence the name e-RHIC. There are different ideas for the electron beams. One involves an energy recovery electron linac installed at one of the RHIC straight sections. The ERL design is quite demanding and there are alternatives possible, which reduce the technical risks. At Jefferson

Lab, the existing 12 GeV electron accelerator would be the electron injector for the Jefferson Lab Electron Ion Collider. A new ion accelerator and a collider would be constructed. The design embraces a figure eight layout for the collider to enhance the spin preservation characteristics and permit the collision of polarized deuterons. At each laboratory the emphasis is on studies and R&D work to mitigate the technical aspects perceived to be high risk. For example, a Jefferson Lab - Institute for Modern Physics (IMP) collaboration recently completed a test of cooling of ion beams using a bunched electron beam at IMP in Lanzhou, China. In addition to risk mitigation value engineering to reduce cost is also being done. This work is being carried out in collaboration with a number of accelerator laboratories and universities in the United States and across the world.

8. **Detector Design**

We are used to the cylindrical symmetry of high energy collider detectors at hadron machines. However historically, we have also tried to exploit the kinematic asymmetries of both the HERA machine with H1 and ZEUS, and the B factories with BaBar and Belle.

The designs evolving for the electron ion collider are attempting to take these considerations a step further. In particular, careful attention is being given to the near beam regions in each direction. The electron direction is usually referred to as the backward direction. A near beam electron detection capability, a tagger, enables low $Q^2$, essentially photoproduction measurements. Such a detector can be used to measure the luminosity by keying on understood electromagnetic processes, and a Compton scattering polarimeter monitors the electron beam polarization.

In the forward, the ion, direction, diffractive processes result in target remnants very close in position and momentum to the unscattered beam. Detection capability in these regions is necessary for studies of coherent interactions. A machine design which provides small beam sizes and large dispersion at the detection point helps considerably. At the same time good acceptance and resolution for momenta much different from that of the beam enables hadron tagging techniques which in turn could provide access to a much broader range of processes, such as effectively-free neutron structure function measurements. Current designs [10] offer a 50-fold improvement in the acceptance of particles with 98% of the beam momentum as compared to ZEUS, from 2% to almost 100%

9. **The NSAC 2015 Long Range Plan and Path Forward**

In the United States, the Nuclear Science Advisory Committee provides guidance for both the Department of Energy and for the National Science Foundation on the development and execution of the nuclear science program. On a regular basis every several years, NSAC develops a long range plan within projected resource constraints. In the 2007 long range plan the concept of an electron collider achieved some prominence. In the 2015 edition of that plan [11], the third recommendation read: "We recommend a high-energy high-luminosity polarized EIC as the highest priority for new facility construction following the completion of the Facility for Rare Isotope Beams (FRIB). "Given the FRIB construction schedule, NSAC is envisaging EIC construction starting in say 2021/2. It is important to note that the DOE has a good record of following the NSAC plans. Indeed as a first step along the path, DOE has commissioned a study by the National Academies of Science to gauge the broader support for an EIC in the scientific community. At the ground level, the recommendation has stimulated a resurgence of user activities. An Electron Ion Collider user group (EICUG) had an inaugural meeting in Berkeley in January, 2016 and followed up with a meeting during the summer at Argonne National Lab. The head count now exceeds 600 collaborators coming from 104 institutions, and, importantly, from 26 different countries.

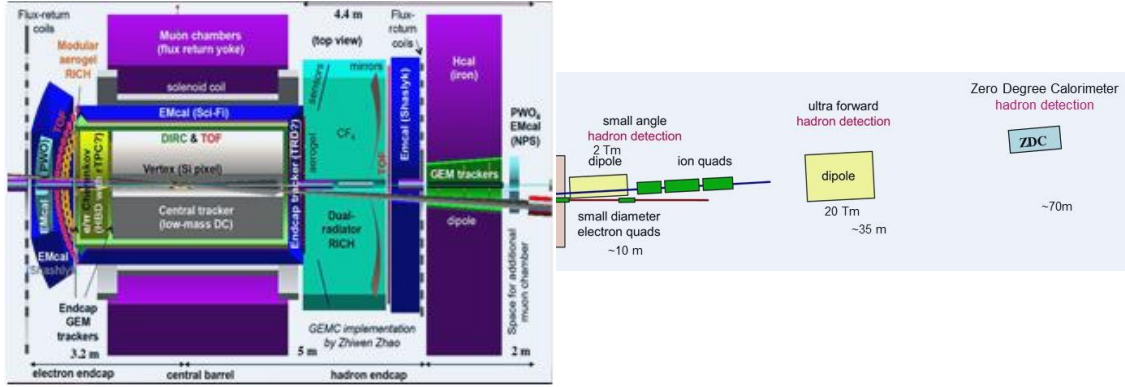

**Figure 5:** Layout of a Jefferson Lab design [10] for an electron ion collider detector; the forward components illustrated to the right, but not to scale, range from 10 m. to 70 m., the latter for the Zero Degree detector.

## 10. Conclusions

We have sketched the motivation for and the concept of an Electron Ion Collider and the physics program possible. This concept has been embraced by the US nuclear physics community and supported by the international nuclear physics community. With its place as a major pillar of the NSAC 2015 Long Range Plan, it is treated seriously by the US Department of Energy. It will be considered during the next year by the broader community of the National Academies of Science. Facility designs exist at the conceptual level and there are collaborative efforts to address technical risk and costs. Detector designs well integrated with the machines are making considerable progress. We can anticipate dramatic improvements compared to the detector coverage enjoyed by the HERA teams. There is palpable excitement and energy being generated by this vision of an Electron Ion Collider.

## 11. Acknowledgements

This paper is based on the work of a large fraction of the international nuclear physics community. In particular, I appreciate input and conversations over the past couple of years from many colleagues including: Markus Diefenthaler, Abhay Deshpande, Rolf Ent, Yulia Furletova, Charles Hyde, Bob McKeown, Pavel Nadel-Turonsky, Rik Yoshida, Zhiwen Zhao. M. B. Stewart was instrumental in the preparation of this manuscript. This work was supported by DOE contract DE-AC05-06OR23177, under which Jefferson Science Associates, LLC, operates the Thomas Jefferson National Accelerator Facility.